\documentclass[preprint]{aastex}

\newcommand{\ie}{{\em i.e.\ }}
\newcommand{\eg}{{\em e.g.},}
\newcommand{\etal}{{\em et al.\ }}
\newcommand{\cf}{{\em cf.\ }}

\newcommand{\UBRI}{{$U, B, R, \&\ I $}}
\newcommand{\UBR}{{$U, B, \&\ R $}}

\newcommand{\U}{{\em U\ }}
\newcommand{\B}{{\em B\ }}
\newcommand{\R}{{\em R\ }}
\newcommand{\I}{{\em I\ }}

\newcommand{\HST}{{\it HST\ }}

\slugcomment{}

\shorttitle{Clustering Evolution}
\shortauthors{Brunner \etal}

\begin{document}

\title{Evolution in the Clustering of Galaxies for $z < 1.0$}

\author{Robert J. Brunner\altaffilmark{1}}
\affil{Department of Astronomy, The California Institute of Technology,
Pasadena, CA 91125}
\email{rb@astro.caltech.edu}

\author{Alex S. Szalay\altaffilmark{1}}
\affil{Department of Physics and Astronomy, The Johns Hopkins University,
Baltimore, MD 21218}
\email{szalay@pha.jhu.edu}

\author{Andrew J. Connolly\altaffilmark{1}}
\affil{Department of Physics and Astronomy, University of Pittsburgh,
Pittsburgh, PA, 15260}
\email{ajc@phyast.pitt.edu}

\altaffiltext{1}{Visiting Astronomer, Kitt Peak National Observatory, National
Optical Astronomy Observatories, which is operated by the
Association of Universities for Research in Astronomy, Inc. (AURA)
under cooperative agreement with the National Science Foundation.}

\begin{abstract}

Measuring the evolution in the clustering of galaxies over a large
redshift range is a challenging problem.
We have developed a new technique which uses photometric redshifts to
measure the angular correlation function in redshift shells. This
novel approach minimizes the galaxy projection effect inherent in
standard angular correlation measurements, and allows for a
measurement of the evolution in the galaxy correlation strength with
redshift. In this paper, we present new results which utilize more
accurate photometric redshifts, which are derived from a multi-band
dataset ($U$,$B$,$R$, and $I$) covering almost two hundred square
arcminutes to $B_{AB} \sim 26.5^m$, to quantify the evolution in the
clustering of galaxies for $z < 1$. We also extend our technique to
incorporate absolute magnitudes, which provides a simultaneous
measurement of the evolution of clustering with both redshift and
intrinsic luminosity. Specifically, we find a gradual decline in the
strength of clustering with redshift out to $z \sim 1$, as predicted
by semi-analytic models of structure formation. Furthermore, we find
that $r_0(z=0) \approx 4.0 h^{-1}$ Mpc for the predictions of linear
theory in an $\Omega_0 = 0.1$ universe.

\end{abstract}


\section{Introduction}

One of the key problems in modern astronomy is understanding how
galaxies form and evolve. Although this problem might seem rather
straightforward, a considerable amount of uncertainty remains,
primarily because of the difficulty in separating out the competing
effects of density evolution, \ie variation in the number of galaxies
with redshift due to clustering or mergers, from luminosity evolution,
\ie the intrinsic evolution in an individual galaxy's spectral energy
distribution.

Historically, two principle techniques have been used to quantify
evolution in the clustering of galaxies. The first technique is to
invert the angular correlation function ($w(\theta)$) using the Limber
equation~\citep{limber54,peebles80} and an observed or model redshift
distribution to estimate the expected change in the amplitude of the
angular correlation function for different magnitude intervals and/or
cosmologies(\eg\ 
~\citealt{koo84,efstathiou91,roche93,ip95,brainerd96}).  The
alternative approach is to compute the spatial correlation function
($\xi(r)$) for different epochs directly using spectroscopic
redshifts~\citep{leFevre96,carlberg97,carlberg99,small99}. These
two techniques, however, suffer from different limitations that have
restricted their utility.

Studies which utilize different magnitude intervals are limited in
that an apparent magnitude selection samples galaxies of different
intrinsic luminosities at different redshifts, complicating the
analysis considerably. Furthermore, the clustering of galaxies on
small scales is the result of a highly non-linear, complex process;
and, therefore, the actual validity of the power law model for the
evolution of the spatial clustering function is not guaranteed,
although it is at least a useful diagnostic. Finally, this approach is
also limited by the applicability of the assumed redshift
distribution, which can heavily influence the theoretical conversion
from angular coordinates to spatial coordinates.

The spectroscopic approach, on the other hand, is hindered either by
the size of the available samples, especially when the data is binned
into distinct redshift intervals, or by the depth or width of the
survey (which implies either a redshift variable, intrinsic luminosity
selection effect or possible contamination from strong
clustering). For example,~\citet{leFevre96} analyze the spatial
clustering for 591 galaxies in the CFRS with $I < 22.5$,
~\citet{small99} analyze the spatial clustering of 831 galaxies with
$r \leq 21^m$ to measure the evolution in the correlation length for
$0.2 \leq z \leq 0.5$, and ~\citet{carlberg97} use a sample of 248
galaxies with $K < 21.5$ while ~\citet{carlberg99} use 2300 high
intrinsic luminosity galaxies distributed over a wide area, to
determine the spatial correlation function for different epochs. 

In this paper, we continue our development of a new technique which
uses photometric redshifts to measure the angular correlation function
in redshift shells~\citep{myThesis,connolly98,brunner99c}. This novel
approach minimizes the galaxy projection effect inherent in standard
angular correlation measurements, while utilizing a significantly
large sample that minimizes the effects of shot noise in our
analysis. By adopting an ensemble approach, we are able to measure the
evolution of clustering with both redshift and intrinsic
luminosity. Unless otherwise noted, we assume $h = 1.0$, $\Omega_M =
0.3$, and $\Omega_\Lambda = 0.7$, throughout this paper.

\section{Data}

The observations and reduction of the data used in this analysis have
been extensively detailed
elsewhere~\citep{myThesis,brunner97,brunner99}. In this section, we
discuss the important points which impact the rest of our discussion.

\subsection{Observations}

The photometric data analyzed in this paper are located at 14:20,
+52:30 covering approximately 0.054 square degrees. All of the
photometric data were obtained using the Prime Focus CCD (PFCCD)
camera on the Mayall 4 meter telescope at Kitt Peak National
Observatory (KPNO). The observations were performed on the nights of
March 31 -- April 3, 1995, March 18 -- 20, 1996, and May 14 -- 16,
1996. The PFCCD uses the T2KB CCD, a $2048^2$ Tektronix CCD with $24
\micron$ pixel scale, which at $f/2.8$ in the 4 meter results in a
scale of $0.47 \arcsec/$pixel and a field of view of $\approx
16.0\arcmin \times 16.0\arcmin$. All observations were made through
the broadband filters: \UBRI.

\subsection{Data Reduction}

The photometric data were reduced in the standard fashion. The data
were photometrically calibrated to the published ~\citet{landolt92}
standard star fields using a curve of growth analysis. A linear
regression on the published stellar magnitude, the instrumental
magnitude, the airmass, and a color term was performed, and the result
translated to a one second standard exposure. We transformed our
magnitude system to the AB system~\citep{okeGunn83} using published
transformations~\citep{fukugita95}.

Source detection and photometry were performed using SExtractor
version 2.0.8~\citep{bertin96} with the appropriate correction for the
background estimation bug applied~\citep{bertin98}. SExtractor was
chosen for its ability to perform matched aperture photometry, using
the same detection image for each program image. Our detection image
was constructed from the \UBRI\ images using an optimal $\chi^2$
process~\citep{szalay98}.

The astrometric solution for our data was determined by matching
against a pre-release version of the \HST Guide Star Catalog
II~\citep{lasker96}. The residuals of the final geometric
transformation to the GSCII for the reference stars were all less than
$0.15$ pixels, or equivalently, less than $0.07\arcsec$.

Our completeness limits were calculated by adding artificially
generated galaxies to the final stacked image. The iterative,
Monte-Carlo approach we used produced a completeness curve, from which
both a $90\%$ and $50\%$ completeness limits in all four bands were
extracted. The $2\%$ and $10\%$ photometric error magnitude limits
were calculated as the mean of all valid detections in the master
catalog which had a measured photometric error that was approximately
the same as the target photometric error ($0.1$ magnitudes for $10\%$
photometry and $0.02$ magnitudes for $2\%$ photometry).

We empirically determined the stellar locus in each band separately
using several complementary techniques: the ratio of a core to total
magnitude, objects which were classified as stars on overlapping \HST
images, and objects which were spectroscopically identified as stars. In
addition, all objects with $I < 20^m$ were visually inspected
and classified as stellar or non-stellar.  The final classification
was constructed by taking the union of the four separate
classifications in each band, resulting in 505 stellar objects. The
number-magnitude distribution of stellar objects agrees with model
predictions~\citep{bahcall80}. The spatial distribution of the stellar
objects is fairly random, with the possible minor exception of the
image corners where the PSF increases due to focal degradations. As a
result, these areas were not utilized during the actual correlation
analysis.

\subsection{Photometric Redshifts}

Many cosmological tests are more sensitive to the sample size (\ie
Poisson Noise) than small errors in distance, which makes them perfect
candidates for utilizing a photometric redshift catalog, including
quantifying the evolution of galaxy clustering. As a result, we have
developed an empirical photometric redshift
technique~\citep{connolly95,brunner97,myThesis,brunner99}, which is
not designed to accurately predict the redshift for a given
galaxy~\citep{baum62} or locate high redshift
objects~\citep{steidel96}. Instead, it is designed to provide distance
indicators which are statistically accurate for the entire sample,
along with corresponding redshift error estimates.

The calibration data for implementing this technique was taken from
overlapping spectroscopic surveys including data from both the Canada
France Redshift survey~\citep{lilly95}, and the Deep Extragalactic
Evolutionary Probe survey~\citep{mould93}. The accuracy of any
empirically derived relationship is predominantly dependent on the
quality of the data used in the analysis --- photometric redshifts
being no exception. As a result, we imposed several restrictions on
the calibrating data in order to minimize the intrinsic dispersion
within the photometric redshift relationship. After imposing our
quality assurance conditions, we were left with 190 galaxies which
formed our calibration sample.

The 190 calibrating redshifts were, therefore, used to derive an
iterative piecewise polynomial fit to the galaxy distribution in the
\UBRI\ flux space. This iterative approach utilizes a global fit to 
determine a rough estimate of the galaxy's redshift, after which a
more accurate local polynomial fit, corresponding to the appropriate
redshift interval, was applied~\citep{brunner99}.  For each derived
polynomial fit, the degrees of freedom remained a substantial fraction
of the original data (a second order fit in four variables requires 15
parameters while a third order fit in four variables requires 35
parameters).

To estimate the error in a photometric redshift for the full
photometric sample, we adopt a bootstrap with replacement algorithm,
in which galaxies are randomly selected from the calibration sample
and, once selected, are not removed from the set of calibrating
galaxies. Thus, at the extremes, one galaxy could be selected 190
consecutive times or, alternatively, each redshift could be selected
exactly once (each of these realizations has the same
probability). This approach is designed to emphasize any
incompleteness in the sampling of the true distribution of galaxies in
the four dimensional space \UBRI\ by the calibration redshifts. In
order to fully account for potential sources of error in the redshift
estimation, the magnitudes of the calibrating sample were drawn from a
Gaussian probability distribution function with mean given by the
measured magnitude and sigma by the magnitude error.

The actual photometric redshift error was calculated from 100
different realizations using this algorithm.  For each different
realization, a photometric redshift was calculated for every galaxy in
the photometric redshift catalog. The actual error was optimally
determined to be given by the normalized difference between the fifth
and second quantiles of the estimated redshift distribution for each
individual galaxy~\citep{brunner99}.  As expected the average
estimated error is the largest at the upper and lower redshift limits
where the incompleteness in the calibrating sample is most
evident. The majority of the rest of the objects with extremely large
redshift errors are blended in one or more bands, which causes these
objects to be isolated from the high density surface delineated by the
majority of galaxies in the four flux space \UBRI. The effect these
objects impart on any subsequent analysis, however, is minimized by
the inclusion of their photometric error, which causes them to be
non-localized in redshift space. As a result, these objects provide a
minimal contribution to many ``redshift bins'' rather than strongly
biasing only a few bins.

A subtle, and often overlooked, effect in any photometric redshift
analysis is the requirement for reliable photometry in all program
filters. Ideally we would obtain accurate redshift estimates for all
galaxies; however, since we need accurate, multi-band photometry in
order to reliably estimate redshifts, we must place restrictions on
the photometric catalog used in the analysis. In particular, we
restrict the full sample of detected sources to those objects which
have both $I_{AB} < 24.0$ and measured magnitude errors $< 0.25$ in
\UBR. This minimizes any selection bias to only faint early-type
galaxies at high redshifts, or very high redshift drop-out objects
(see~\citealt{brunner99} for more discussion), neither of which
significantly affect the rest of our analysis.

\section{Ensemble Approach}

Due to the lower precision of photometric redshift determination as
compared to spectroscopic redshifts (roughly a factor of 20), we have
developed a new, statistical approach to quantifying the evolution of
galaxies~\citep{myThesis,connolly98,brunner99}. This approach
capitalizes on our ability to reliably generate not only a redshift
estimate for a galaxy using broadband photometry, but also a reliable
redshift error estimate. As a result, we define the probability
density function, $P(z)$, for an individual galaxy's redshift to be a
Gaussian probability distribution function with mean ($\mu$) given by
the estimated photometric redshift and standard deviation ($\sigma$)
defined by the estimated error in the photometric redshift.

\[
P(z) = \frac{1}{\sigma \sqrt{2\pi}} e^{\left(-\frac{(z -
\mu)^2}{2\sigma^2}\right)}
\]

In order to measure an interesting cosmological quantity, we generate
multiple ensembles (or realizations) of the relevant properties of the
galaxy distribution using the statistical redshift and redshift error
estimates (see~\citealt{brunner99} for an application of this
technique to the number-redshift distribution). The quantity of
interest is determined as the mean of the multiple realizations, and
the associated error is given by the corresponding standard deviation.

In order to divide our sample by intrinsic luminosity, we determined
the absolute magnitude distribution of the galaxies in our catalog in
an ensemble approach. First, we created different realizations of our
galaxy catalog. In order to minimize any systematic errors, we
selected the apparent magnitude of each galaxy from a Gaussian
probability distribution function with mean and sigma given from the
original photometric catalog measurements. Similarly, the redshift of
each galaxy was drawn from a separate Gaussian probability
distribution function with mean and sigma given from the photometric
redshift and corresponding redshift error estimate. The $k-$correction
was determined using the spectral classification which was part of the
original redshift estimation procedure. For galaxies with large
photometric redshifts, occasional discordant redshifts were calculated
(\ie outside the range of our calibration sample --- $z < 0$, or $z >
1.2$) in which case the galaxy was dropped from that particular
realization.

Together, these quantities were used to determine the absolute
magnitude for each galaxy in 100 different ensemble distributions. The
absolute magnitude for each galaxy was calculated as the mean over the
different realizations, appropriately normalized to account for
possible discordant redshifts as discussed above. The resultant
distributions for the $U$ and $B$ bands are displayed in
Figure~\ref{abs-mag}.

\section{Analysis}

Before computing the angular correlation function, we quantified our
efficiency in detecting galaxies as a function of pixel location.  The
primary areas where this effect is important are around bright stars,
in charge transfer trails, and near the edge of the frame due to edge
effects or focus degradations. We, therefore, defined bounding boxes,
for each of the four stacked images, which contained all of the
observable flux for the saturated stars within the image. In the end,
a total of 15 regions were masked out in the \U frame, 17 regions were
masked out in the \B frame, 45 regions were masked out in the \R
frame, and 36 regions were masked out in the \I frame. We also masked
both the edge and corners of each frame in order to reduce the effects
of PSF variations on our object detection efficiency. These four mask
files were concatenated to produce a total mask file which was used
for the calculation of the angular correlation function in different
redshift or absolute magnitude intervals.

We used the optimal estimator~\citet{landy93} $(DD - 2DR + RR)/RR$,
where $D$ stands for data and $R$ stands for random, to determine the
angular correlation function. This required counting the number of
observed pairs (that were not within masked areas), which was done in
10 bins of constant width $\Delta\lg(\theta) = 0.25$, centered at
$\theta = 4.3\arcsec$, to $\theta = 759.6\arcsec$. One thousand
objects were randomly placed in the non masked areas within the image,
and the data-random (DR) and random-random (RR) correlation functions
were calculated for the same angular bins used for the data-data (DD)
auto-correlation function. Before applying the estimator, each of the
correlation measurements were scaled by the appropriate number of
pairs.  This process was repeated ten separate times, and the results
averaged to minimize any systematic effects.

This estimator uses the calculated number density of galaxies within
the CCD frame to estimate the true mean density of galaxies. The small
angular area of our images introduces a bias in the estimate, commonly
referred to as the ``Integral Constraint''~\citep{peebles80}. We
estimated a correction for this bias following the prescription
of~\citet{landy93}, which is subtracted from the estimated value.

The error in the estimation of the angular correlation function was
assumed to be Poisson in nature. As a result, we calculated the error
in the angular correlation estimation as the square root of the number
of random-random pairs in each angular bin, scaled by the relevant
number of data points.

The angular correlation function is generally parameterized in the following fashion:
\[w(\theta) = A_w \theta^{-\delta}\]
where the exponent has previously been shown to be $\delta = 0.8$. In
general, the amplitude of the correlation function ($A_w$) is
calculated (assuming the previous value for $\delta$) by minimizing
the $\chi^2$ with respect to $A_{w}$~\citep{press92}:

\[
A_{w} = \left(\sum_{i = 1}^{N} \lg(w(\theta)_i)  + 0.8 \sum_{i = 1}^{N}
\lg(\theta_i) \right) / N
\]
This calculation can be significantly affected by outliers; and, as a
result, we adopt the more robust technique of minimization of the
absolute deviations to determine the angular correlation
amplitude. For our simple case, this technique reduces to finding the
median of the amplitudes at each angle~\citep{press92}.

\subsection{The Angular Correlation Function: $w(\theta)$}

Although not the primary aim of this paper, we measured the angular
correlation function in different apparent magnitude intervals to
compare with previous work. For each program filter, the absolute
upper magnitude limit for the data used in the estimation of the
angular correlation function was set at the $90\%$ catalog
completeness limit and the lower magnitude limit was always set to
$15^m$. The angular correlation function was then determined in four
different magnitude ranges (each offset from the next by $0.5^m$). The
upper magnitude limits, and the corresponding number of objects are
listed in Table~\ref{angCorSample}. Each estimation was repeated 100
times. 

The amplitude of the different correlation functions for each band at
$\theta = 1.0\arcsec$ are listed in Table~\ref{angCorAmp}. For
brevity, we only display the results for the B band. Thus, in
Figure~\ref{bAngCor} the actual correlation measurements for the
different magnitude bins are displayed, while in Figure~\ref{awb} the
measured correlation amplitudes are compared to comparable published
results (\eg\  ~\citealt{koo84,roche93,ip95}), showing remarkably good
agreement. The error bars were calculated from the one sigma upper and
lower measurements of the amplitude of the angular correlation
function.

\subsection{The Multi-Variate Angular Correlation Function: $w(\theta, z)$}

Using the 3052 objects in the photometric redshift--template SED
catalog, the multivariate angular correlation function $w(\theta,
z_{P})$ was determined for four different redshifts by binning the
data in non-overlapping redshift bins of width $\Delta z_{P} = 0.2$
centered at $z_{P} = 0.4$ to $z_{P} = 1.0$. The four different
functions were calculated for the 445, 573, 946, and 582 objects in
the different respective redshift bins. We display, in
Figure~\ref{awz}, the measured change in the amplitude of the angular
correlation function with redshift (\ie $A_{w}$), and hence the
strength of the angular correlation function at a fixed angular
separation. The error in $A_{w}(z)$ was calculated by estimating the
amplitude in each redshift interval for the one sigma upper and lower
values.

In order to compare this result with theoretical expectations (\eg\
semi-analytic theory), which are determined in spatial coordinates, it
is necessary to convert between angular correlation measurements and
spatial correlation quantities. The standard technique for determining
this transformation is to assume a power law model for the spatial
clustering~\citep{peebles80,leFevre96},
\[
\xi(r,z) 
	= \left( \frac{r}{r_{0}(z)} \right )^{-\gamma}
	= \left( \frac{r}{r_{0}(0)} \right )^{-\gamma} (1 + z)^{-(3 + \epsilon)}
\]
where $\epsilon$ represents a parameterization of the evolution of the
spatial correlation function, and the correlation length is measured
in physical units. The conversion, for small angular separations,
between angular and spatial coordinates is accomplished via the
relativistic form of Limber's
equation~\citep{limber54,peebles80}. Specifically, this results in the
following conversion for the amplitude of the angular correlation
function within the redshift region of interest:
\begin{equation}
A_w
	= 	\left( 
		\frac {\int_0^{\infty} G(z) B(\gamma) (\frac{dN}{dz})^2 dz}
		{\left[ \int_0^{\infty} \frac{dN}{dz} dz \right]^2}
		\right) r_0^\gamma
\label{limberE}
\end{equation}
where,
\[
G(z) = (1 + z)^{-(3 + \epsilon)} \sqrt{1 + \Omega_0\ z}\ x(z)^{(1-\gamma)},
\]
\[
B(\gamma) 
	= 	\left(\frac{3600.0 * 180.0}{\pi}\right)^{(\gamma - 1)}
		\left(\frac{H_0}{c}\right)^\gamma
		\frac{\Gamma(\frac{1}{2}) \Gamma(\frac{(\gamma - 1)}{2})}
			{\Gamma({\frac{\gamma}{2}})}		
\]
is a constant quantity,
\[
x(z) 
	= 2 \frac{((\Omega_0 - 2) (\sqrt{1 + \Omega_0 z} + 2 - \Omega_0 + \Omega_0 z)}
	{\Omega_0^2 (1 + z)^2},
\]
is the angular diameter distance~\citep{weinberg72}, and $dN/dz$ is
the number of galaxies per unit redshift. Local spectroscopic surveys
have determined that $\gamma \approx 1.8$ and $r_{0} \sim 5.0$ Mpc
in co-moving coordinates~\citep{carlberg99}.

By assuming a uniform redshift distribution, we have calculated the
tracks of different evolutionary models for $\Omega_0 = 1.0$ and
$\Omega_0 = 0.1$, which are displayed, along with the measured values
of the amplitude of the angular correlation function in
Figure~\ref{awz}. Of the three different scenarios, fixed clustering
in co-moving coordinates ($\epsilon = -1.2$) are the least consistent
with our data, independent of the value of $\Omega_0$. The results for
clustering fixed in proper coordinates ($\epsilon = 0.0$) are good,
independent of the value of $\Omega_0$, while the predictions of
linear theory ($\epsilon = 0.8$) are in agreement for higher values of
$\Omega_0$.

While this result is interesting, we can take an additional step in
order to directly compare our measurements with published results from
spectroscopic surveys. Using Equation~\ref{limberE} and an ensemble
averaged empirical redshift distribution, we can actually transform
our measurement of the angular correlation amplitude into a
determination of the spatial correlation scale length within a given
redshift interval, since we are able to empirically compute our
observed redshift distribution (see~\citealt{brunner99} for more
details). Essentially, this only involves adding a top-hat window
function (corresponding to the appropriate redshift interval) to the
integrands in Equation~\ref{limberE}, since the transformation will be
applied individually to each redshift bin in which we calculated the
angular correlation amplitude.

The process is complicated, however, by the fact that our $dN/dz$ is
determined using photometric redshifts, while the transformation
assumes a spectroscopic redshift interval. From Figure~\ref{zs-zp},
although small, the dispersion between spectroscopic redshifts and
photometric redshifts is not zero. The conversion between a
spectroscopic interval and a photometric redshift interval accounts to
a broadening of the top-hat function, which we accomplish using two
Gaussians centered on the endpoints of the redshift interval, so that
the new window function is given by
\[
W(z) = 
\left\{	
	\begin{array}{ll}
		e^{\frac{-(z - z_1)^2}{\sigma^2}} 	& 0 \leq z < z_1 \\
		1 					& z_1 \leq z \leq z_2 \\
		e^{\frac{-(z - z_2)^2}{\sigma^2}} 	& z_1 < z < \infty
\end{array}
\right.,
\]
where the desired redshift interval is given by $z \in [z_1,z_2]$, and
$\sigma = 0.061$ is the measured dispersion in the photometric
redshift relation. 

The results of the transformation determine the correlation length
within the given redshift bin (\ie $r_0(z)$), and are displayed in
Figure~\ref{roz} and also tabulated in Table~\ref{r0zt} for $\Omega_0
= 0.1$, and the three canonical values of $\epsilon$.  In addition,
the values of the correlation length extrapolated to $z = 0$ (\ie
$r_0(0)$) are tabulated in Table~\ref{r0} for three different values
of $\Omega_0$. The strong dependence of the correlation scale length
on the value of the evolutionary parameter ($\epsilon$) is a direct
result of the Cosmological term ($G(z)$) in
Equation~\ref{limberE}. Relative to the predictions of linear theory
($\epsilon = 0.8$), the results for fixed clustering in co-moving
coordinates ($\epsilon = -1.2$), are suppressed by an additional
factor of $\sim (1 + z)$, or roughly a factor of $1.5$--$2$ at the
redshifts of interest (recall that $r_0(z) \propto G(z)^{-\gamma}$).

When comparing our results to previous spectroscopic results, we
clearly show excellent agreement with recent
measurements~\citep{carlberg99,small99} when the predictions of linear
theory are used to quantify the evolution of clustering. This is
extremely encouraging for our technique, as we uniformly sample a
larger redshift range, showing, for the first time within the same
dataset, the slight decrease in the correlation strength for $z < 1$,
as predicted by semi-analytic models of galaxy
formation~\citep{baugh99}. On the other hand, the measurement of the
correlation scale length from the CFRS data~\citep{leFevre96} are only
in agreement with our results for fixed clustering in co-moving
coordinates, which disagrees with the hierarchical growth of dark
matter halos~\citep{baugh99}. The discrepancy between the CFRS and
other measurements is most likely due to their relatively small sample
size, their small fields, and their neglect of the redshift evolution
of the Luminosity function~\citep{carlberg99}.

\subsection{The Multi-Variate Angular Correlation Function: $w(\theta, z, M)$}

While important, the evolution of the angular correlation function
with redshift smoothes over the galaxy luminosity function, ignoring
variations in clustering between galaxies of different intrinsic
luminosity. As a result, we subdivided our sample into three redshift
intervals ($0.2 \leq z \leq 0.6$, $0.4 \leq z \leq 0.8$, $0.6 \leq z
\leq 1.0$), and measured the angular correlation function as a
function of both $U$ and $B$ absolute magnitude in intervals of
$2.0^m$, from $22^m$ to $12^m$. In order to improve the number of
galaxies in the faint end of our analysis, we rebinned the data so
that the faint bin was four magnitudes wide (\ie $16^m$--$12^m$). In
the end, we obtained twelve different measurements of the multivariate
angular correlation function ($w(\theta, z, M)$) as a function of both
$U$ and $B$ absolute magnitudes.

The number of subdivisions used in this particular analysis
reintroduced one of the problems our new technique was designed to
avoid, namely the effects of small sample size. We, therefore, only
used a joint redshift-absolute magnitude bin when the number of
objects in the bin exceeded one hundred galaxies. From
Figures~\ref{umz},~\ref{bmz} ($U$ and $B$ respectively), it is clear
that there is no obvious evolution within a given redshift interval
with intrinsic luminosity (although this could be a result of too few
galaxies). Between redshift intervals, however, there is strong
evolution in the amplitude of the angular correlation function
($A_w$), which, given the wider redshift bins, is completely
consistent with the results of the previous section.

In order to quantify this evolution, we fit a line of zero slope to
the points (\ie a mean value for each redshift interval). Between the
first two redshift bins (roughly $z \approx 0.4$ and $z \approx 0.6$),
$A_w$ drops by approximately a factor of two. For the $U$ band
measurement, the drop between the second and third redshift intervals
(roughly $z \approx 0.6$ and $z \approx 0.8$) is approximately $20\%$,
while the $B$ band shows another factor of approximately two
decline. This result is not surprising in the context of hierarchical
structure formation where one expects objects to be more strongly
cluster with decreasing redshift for $z < 1$ (\eg
~\citealt{kauffmann99}).

Unfortunately, we do not have enough galaxies to definitely test for
clustering evolution with redshift for a given population with a fixed
intrinsic luminosity. On the other hand, our results do seem to
indicate that the evolution is not strongly dependent on intrinsic
luminosity as there appears to be no clear evidence for variation in
the clustering amplitude within a given redshift interval.  The
difference in the drop in the amplitude between the middle and high
redshift intervals is most likely due to the lower, average intrinsic
luminosity of the galaxies in the $B$ band as compared to the $U$
band. These points will need to be addressed with future datasets.

\section{Conclusions}

In this paper, we have presented several calculations of the angular
correlation function, as a function of different apparent magnitude
intervals, as a function of redshift, and as a function of both
redshift and absolute magnitude. The technique we have demonstrated is
less sensitive to redshift distortions than the spatial correlation
approach due to the width of our redshift bins. Furthermore, our
technique does not require model predictions for the redshift
distribution of galaxies as does the apparent magnitude interval
approach. Future work in this area will incorporate spectroscopic
redshifts into the calculation in order to provide limited distance
information (\cf~\citealt{phillipps85}), as well as witness the
application of these techniques to larger surveys.

While not the main point of this paper, the variation of the amplitude
of the angular correlation with apparent magnitude is in good
agreement with previously published results, which strengthens the
rest of our analysis. Furthermore, we demonstrated, for the first time
from within a single dataset, the slight evolution in both the
amplitude of the angular correlation function, and the spatial
correlation scale length with redshift for $z < 1$, as predicted by
semi-analytic models of structure
formation~\citep{baugh99,kauffmann99}. These results suggest low
values for $\Omega_0$, and allow either fixed clustering in proper
coordinates, or the predictions of linear theory.

Finally, we measured the evolution of the amplitude of the angular
correlation function with both redshift and intrinsic luminosity. The
amplitude of the angular correlation function drops dramatically with
redshift. Interestingly enough, however, we do not see significant
variation in the strength of clustering within a given redshift
interval as a function of intrinsic luminosity. This type of evolution
might be naively expected if the luminosity of a galaxy uniquely
mapped to the mass of the dark matter halo at the given redshift in
which it resides (\ie more luminous galaxies cluster more strongly
within a given redshift interval). Most likely, either we have too
small of a sample to place significant limits on the variation of
clustering with luminosity, or else the relatively constant clustering
amplitude as a function of luminosity is indicative of luminosity
evolution complicating the analysis.

Future surveys, both photometric and spectroscopic (\eg\ SDSS) will
provide extremely useful datasets with which we can explore these
ideas in greater detail. In the near future, this area will witness a
merging of observations, semi-analytic theory, and N-body simulations,
finally providing hope that we will be able to unambiguously quantify
the clustering evolution of galaxies.

\acknowledgments

First we wish to acknowledge Gyula Szokoly for assistance in obtaining
the data. We also would like to thank Barry Lasker, Gretchen Greene,
and Brian McLean for allowing us access to an early version of the GSC
II. We also wish to acknowledge useful discussions with Pat Cote, Rich
Kron, Lori Lubin, and Ray Weymann. We thank the anonymous referee for
valuable suggestions on improving this work.  This research has made
use of NASA's Astrophysical Data System Abstract Service. AS
acknowledges support from NASA LTSA (NAG53503) and \HST Grant
(GO-07817-04-96A), AJC acknowledges partial support from \HST
(GO-07817-02-96A) and LTSA (NRA-98-03-LTSA-039).

\newpage

\figcaption[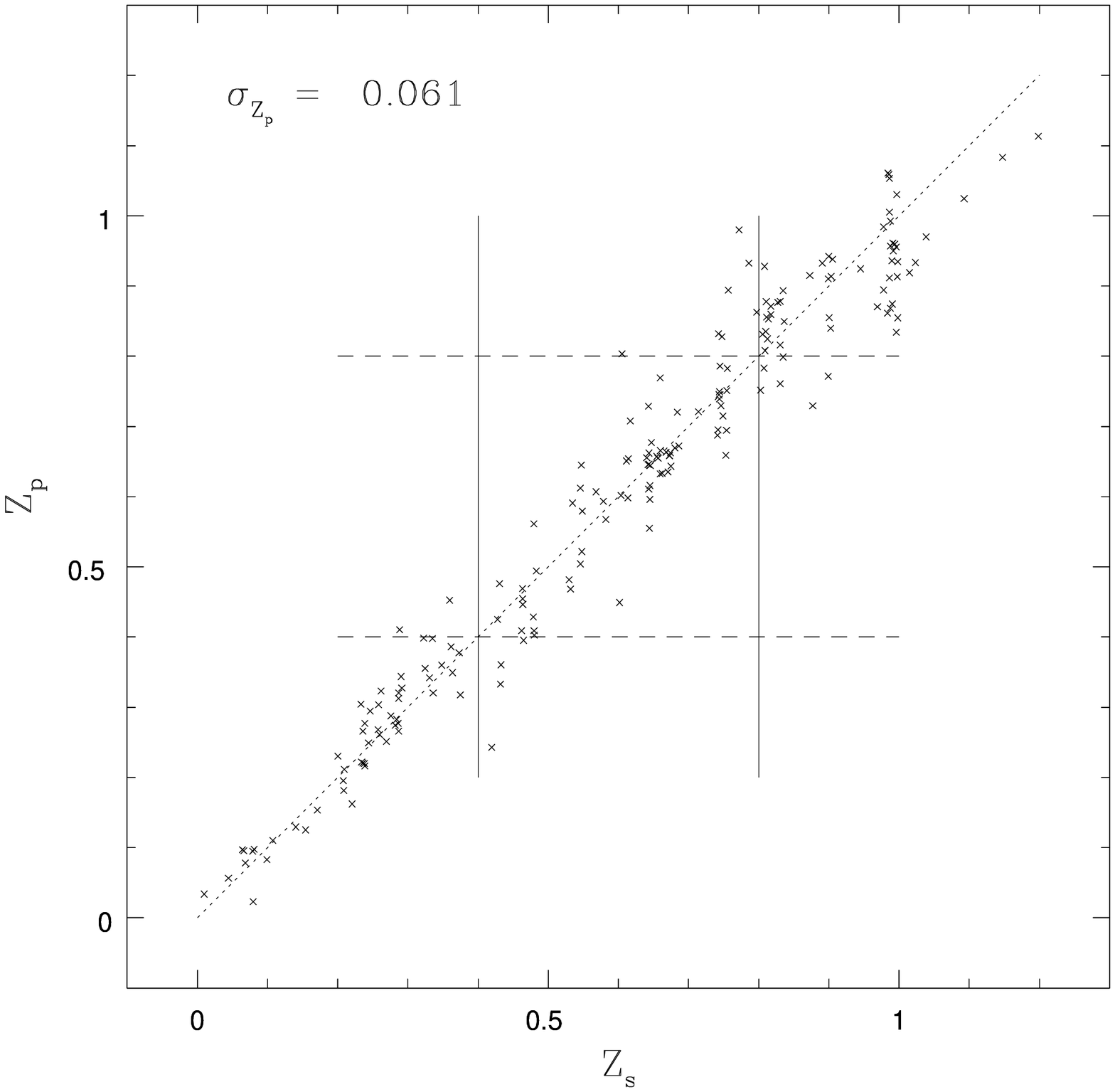]{The dispersion in the empirical photometric redshift relation 
used in the analysis ($\sigma = 0.061$). The short solid and dashed
lines are used to compare the transformation between spectroscopic
redshift intervals and photometric redshift intervals.
\label{zs-zp}}

\figcaption[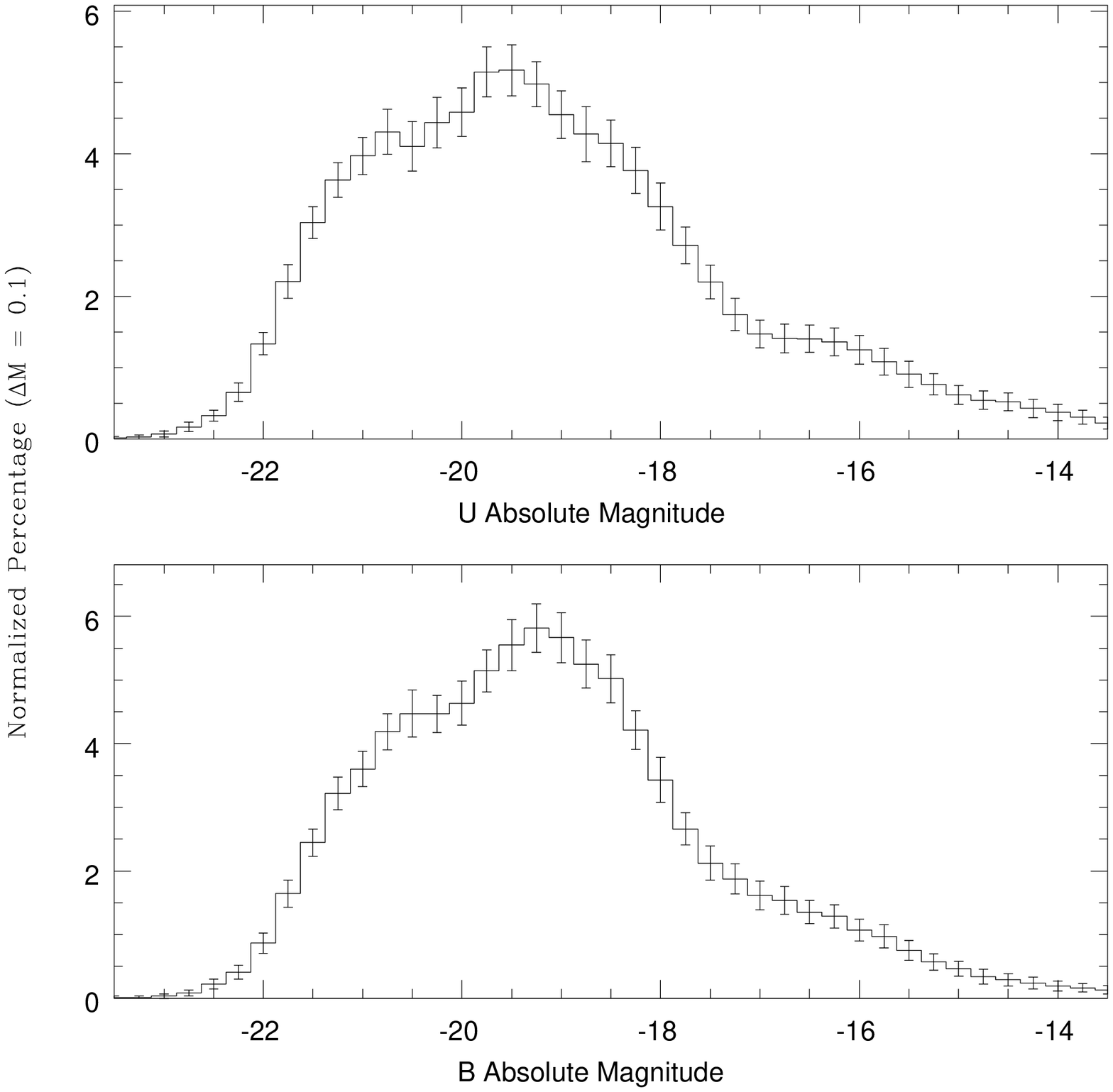]{The absolute magnitude distribution of the photometric galaxy 
sample in the U (top) and B (bottom) bands. The determination of the
distributions used the ensemble approach (see text for more details).
\label{abs-mag}}

\figcaption[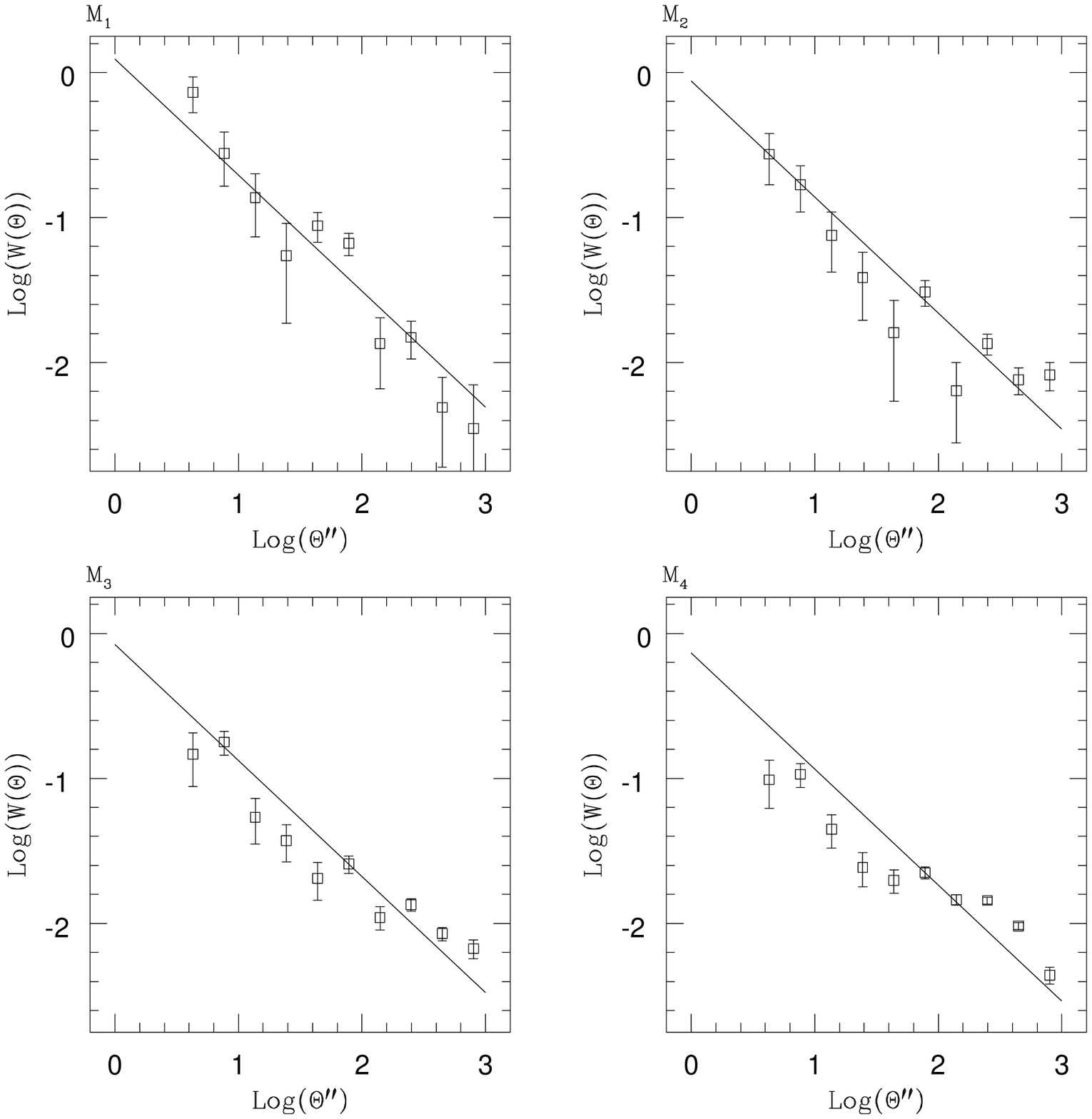]{The \B band angular correlation function for all spectral
types. The solid line is a fit to the data of a line with fixed slope
$\delta = -0.8$ using the method of minimization of the absolute
deviations~\cite{press92}.
\label{bAngCor}}

\figcaption[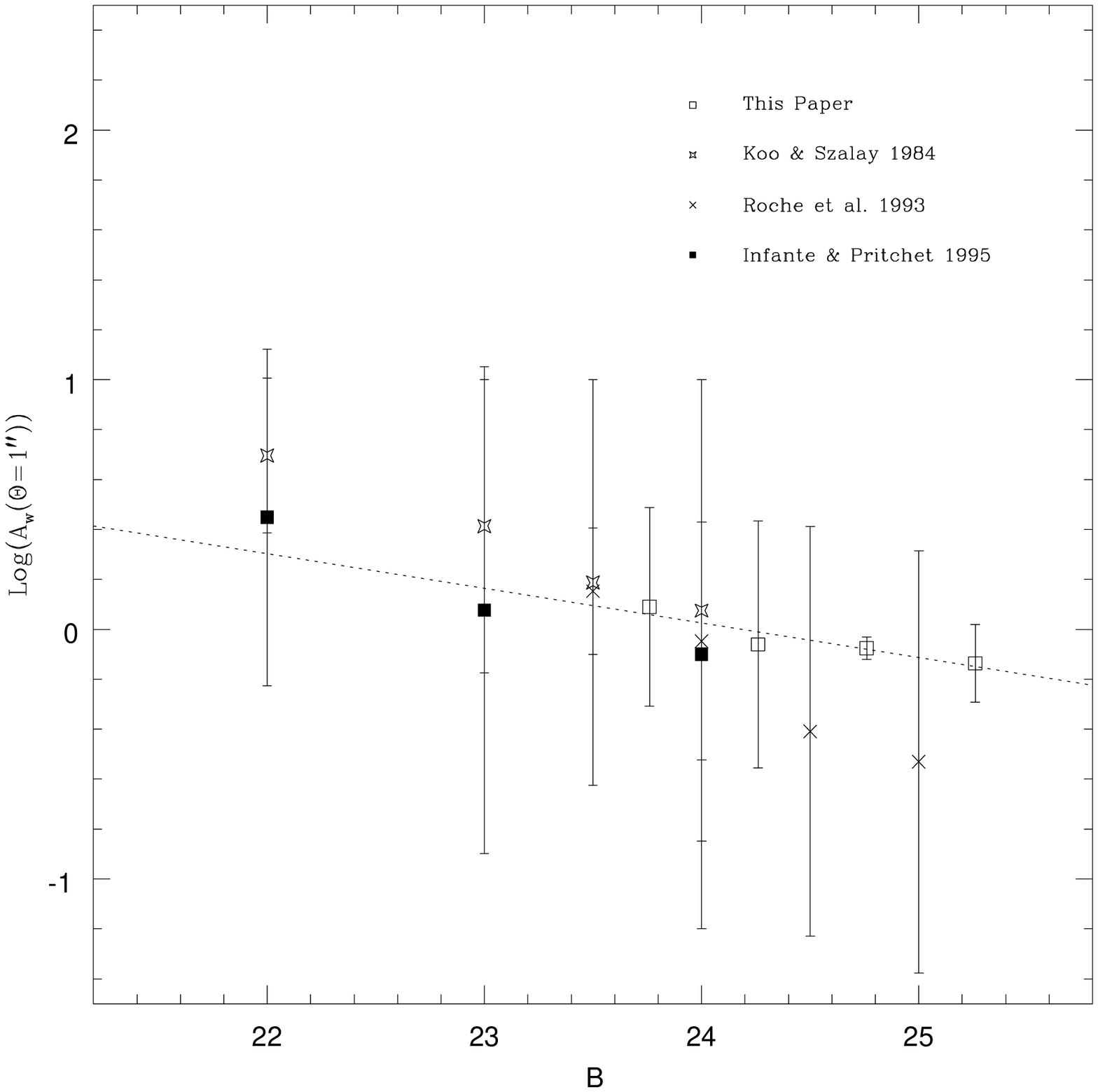]{The variation in the amplitude of the \B band angular 
correlation function for the data presented in this paper, as well as
other similar published results. The dashed line is a least squares
fit to our data points, and is included merely as a visual aid. This
demonstrates the excellent agreement between both our data and
technique with previous results.
\label{awb}}

\figcaption[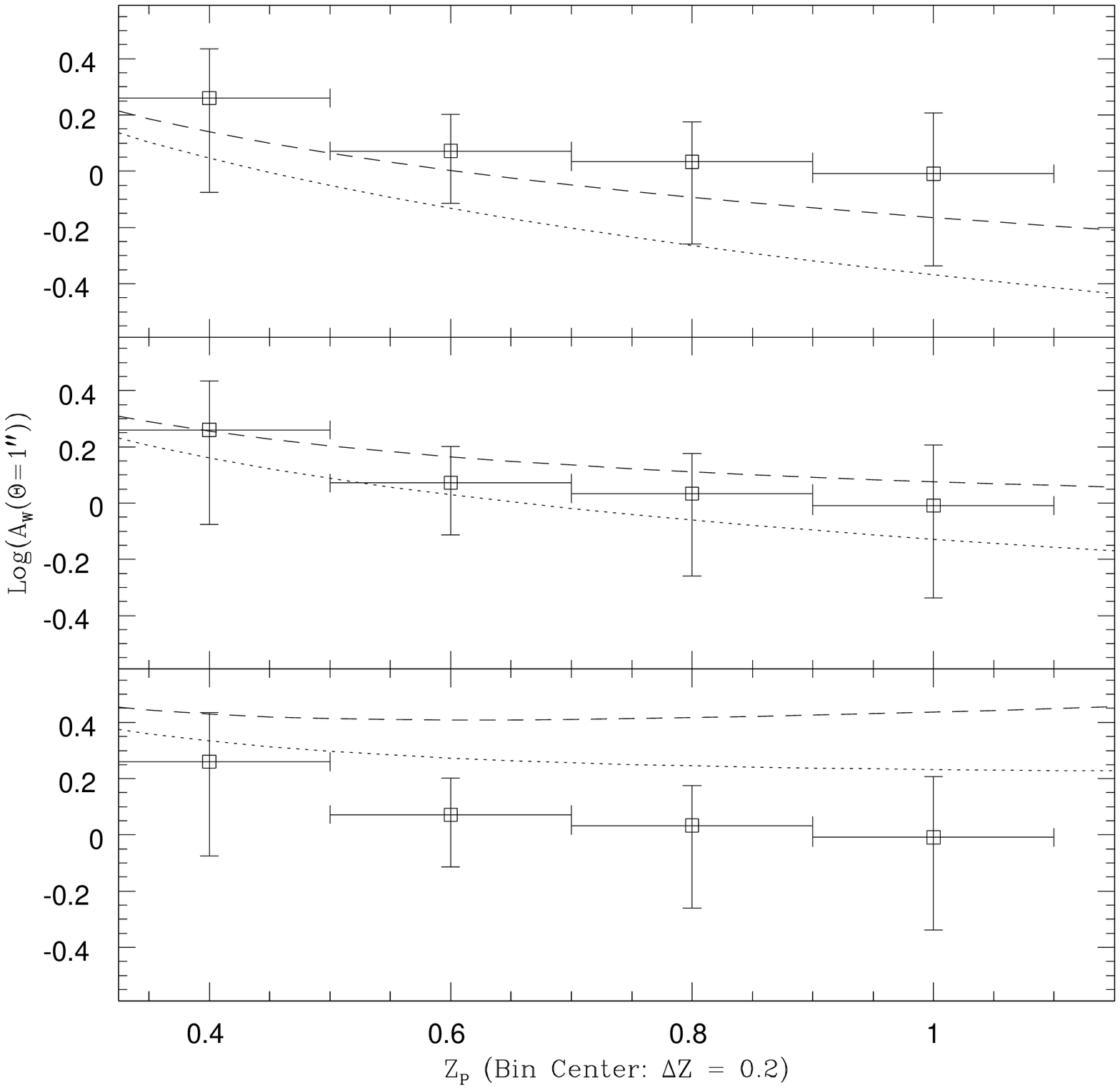]{The evolution in the amplitude of the angular correlation
function with redshift. The two lines are predictions for $\Omega_0 =
0.1$ (dotted line) and $\Omega_0 = 1.0$ (dashed line) using Limber's
equation~\citep{peebles80}. The top panel assumes the evolution
parameter derived from linear theory ($\epsilon\ = 0.8$), the middle
panel assumes fixed clustering in proper coordinates ($\epsilon\ =
0.0$), and the bottom panel assumes fixed clustering in co-moving
coordinates ($\epsilon\ = -1.2$).
\label{awz}}

\figcaption[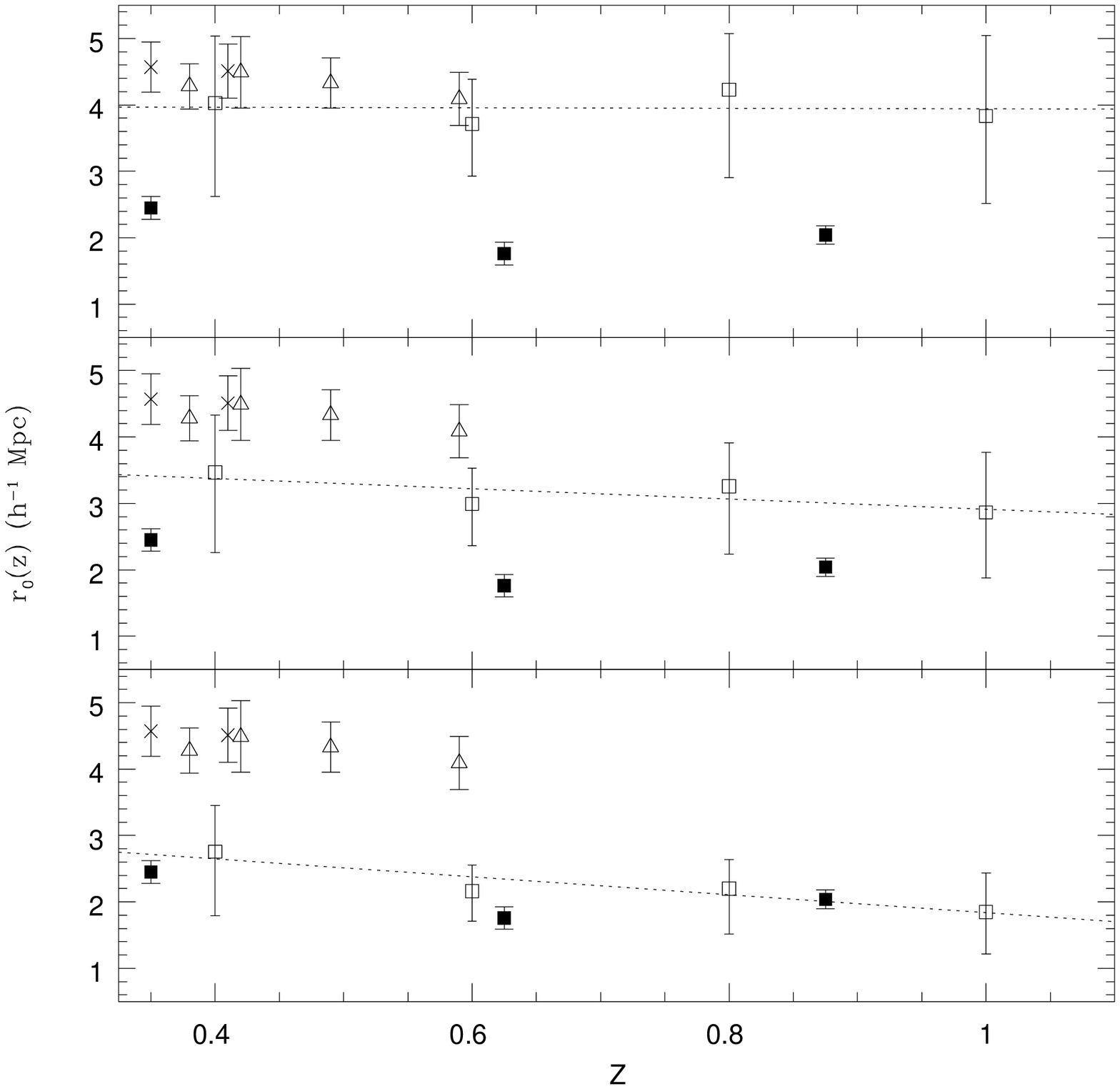]{The evolution in the spatial correlation scale length ($r_0$) 
with redshift, assuming $\Omega_0 = 0.1$ The data presented in this
paper are indicated by open squares ($\square$). The different panels
correspond to the three different canonical values for the
parameterization of the evolution in the spatial correlation function:
upper panel, predictions of linear theory ($\epsilon = 0.8$), middle
panel, clustering fixed in proper coordinates ($\epsilon = 0.0$),
bottom panel, clustering fixed in co-moving coordinates ($\epsilon =
-1.2$). Overplotted are spectroscopic survey measurements with error
bars (transformed to our Cosmology using the prescription
of~\citealt{leFevre96}) of the correlation scale length:
($\times$)~\citealt{small99}, ($\bigtriangleup$)~\citealt{carlberg99},
and ($\blacksquare$)~\citealt{leFevre96}. Although not shown (in order
to improve the clarity of the figure), each of our data points has an
uncertainty in the horizontal direction of $\pm 0.1$.
\label{roz}}

\figcaption[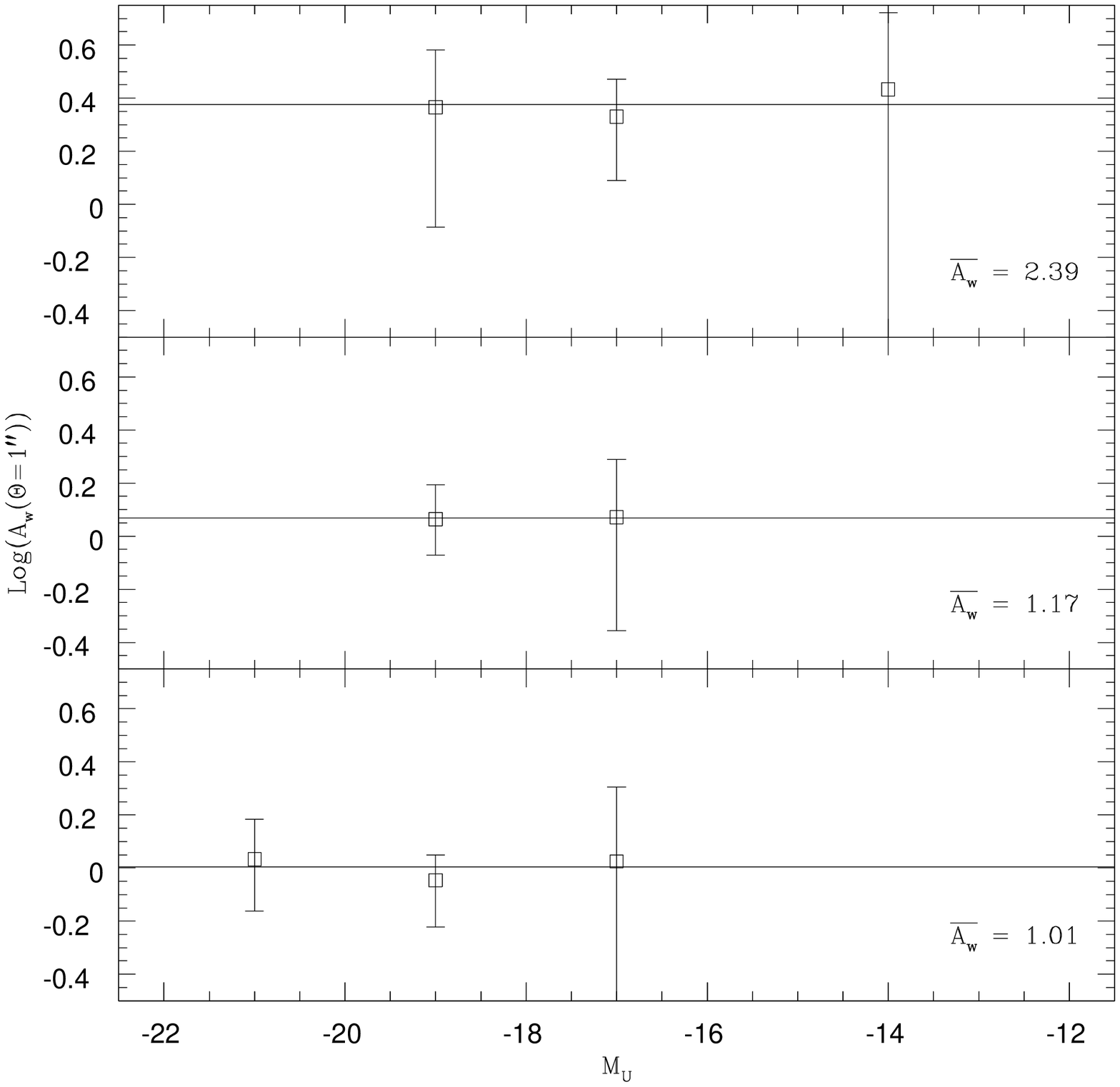]{The evolution in the correlation scale length with 
both redshift and absolute $U$ magnitude. The three panels correspond
to the three different redshift intervals utilized: $0.2 \leq z \leq
0.6$ (top panel), $0.4 \leq z \leq 0.8$ (middle panel), and $0.6 \leq
z \leq 1.0$ (bottom panel). The absolute magnitude ordinal is assigned
as the median of the appropriate absolute magnitude bin. As a result,
the error in absolute magnitude is $\pm 1.0^m$ for the first three
magnitude bins, and $\pm 2.0^m$ for the last magnitude bin. In each
bin, we determine the mean value for the correlation amplitude, which
declines with decreasing redshift.
\label{umz}}

\newpage

\figcaption[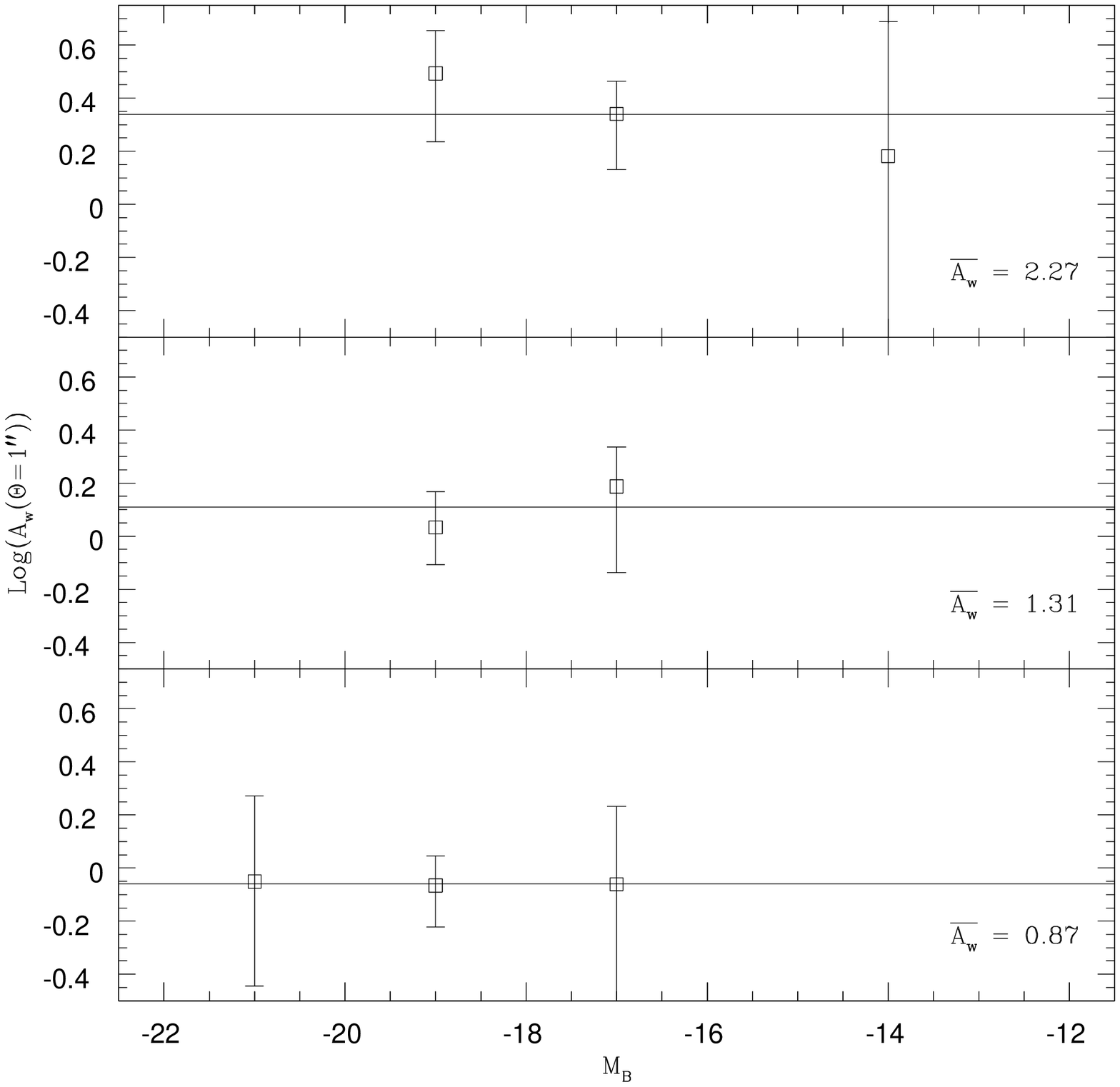]{The evolution in the correlation scale length with 
both redshift and absolute $B$ magnitude. The three panels correspond
to the three different redshift intervals utilized: $0.2 \leq z \leq
0.6$ (top panel), $0.4 \leq z \leq 0.8$ (middle panel), and $0.6 \leq
z \leq 1.0$ (bottom panel). The absolute magnitude ordinal is assigned
as the median of the appropriate absolute magnitude bin. As a result,
the error in absolute magnitude is $\pm 1.0^m$ for the first three
magnitude bins, and $\pm 2.0^m$ for the last magnitude bin. In each
bin, we determine the mean value for the correlation amplitude, which
declines with decreasing redshift.
\label{bmz}}

\newpage

\begin{deluxetable}{ccccc}
\tablecaption{The upper magnitude limit and corresponding number of objects
used in the estimation of the angular correlation function for each
band.
\label{angCorSample}}
\tablewidth{0pt}
\tablehead{
\colhead{Band} 
&\colhead{$M_{1}$}
&\colhead{$M_{2}$}
&\colhead{$M_{3}$}
&\colhead{$M_{4}$}
}
\startdata
U	& 24.44 (1266)	& 24.94 (2257)	& 25.44 (3802)	& 25.94 (5668)	\\
B	& 23.76 (812)	& 24.26 (1524)	& 24.76 (2685)	& 25.26 (4365)	\\
R	& 22.95 (1003)	& 23.45 (1647)	& 23.95 (2673)	& 24.45 (4035)	\\
I	& 22.47 (1002)	& 22.97 (1579)	& 23.47 (2429)	& 23.97 (3602)	\\
\enddata
\end{deluxetable}

\begin{deluxetable}{ccccc}
\tablecaption{The amplitude $A_{w}$ of the angular correlation
function at $\theta = 1.0\arcsec$ assuming a relation $w(\theta) =
A_{w}\theta^{-0.8}$ for the different magnitude intervals.
\label{angCorAmp}}
\tablewidth{0pt}
\tablehead{
\colhead{Band} 
&\colhead{$M_{1}$}
&\colhead{$M_{2}$}
&\colhead{$M_{3}$}
&\colhead{$M_{4}$}
}
\startdata
U	&0.81	&0.70	&0.69	&0.60	\\
B	&1.23	&0.87	&0.84	&0.73	\\
R	&1.80	&1.26	&1.03	&0.94	\\
I	&1.87	&1.42	&1.08	&0.94	\\
\enddata
\end{deluxetable}

\begin{deluxetable}{cccc}
\tablecaption{The measured correlation scale length and corresponding 
error at different redshifts for $\Omega_0 = 0.1$ and different values
of the evolutionary parameters ($\epsilon$).
\label{r0zt}}
\tablewidth{0pt}
\tablehead{
\colhead{} &
\multicolumn{3}{c}{$r_0(z)$}
\\
\cline{2-4}
\colhead{$z$} &
\colhead{$\epsilon = 0.8$} & 
\colhead{$\epsilon = 0.0$} & 
\colhead{$\epsilon = -1.2$}
}
\startdata
0.4	&	4.03 (+1.01/-1.41) & 3.47 (+0.87/-1.21) & 2.76 (+0.69/-0.96)	\\
0.6	&	3.71 (+0.67/-0.78) & 2.99 (+0.54/-0.63) & 2.16 (+0.39/-0.46)	\\
0.8	&	4.23 (+0.85/-1.32) & 3.26 (+0.65/-1.02) & 2.20 (+0.44/-0.69)	\\
1.0	&	3.83 (+1.22/-1.31) & 2.86 (+0.91/-0.98) & 1.85 (+0.59/-0.63)	\\
\enddata
\end{deluxetable}

\begin{deluxetable}{cccc}
\tablecaption{The correlation scale length and associated errors extrapolated to 
$z = 0.0$ for different values of Cosmological and Evolutionary
parameters.
\label{r0}}
\tablewidth{0pt}
\tablehead{
\colhead{} &
\multicolumn{3}{c}{$r_0(z = 0)$}
\\
\cline{2-4}
\colhead{$\Omega_0$} &
\colhead{$\epsilon = 0.8$} & 
\colhead{$\epsilon = 0.0$} & 
\colhead{$\epsilon = -1.2$}
}
\startdata
0.1	&3.98 (+0.65/-1.12)&3.69 (+0.65/-1.07)&3.19 (+0.61/-0.95)\\
0.3	&3.94 (+0.66/-1.11)&3.63 (+0.66/-1.05)&3.12 (+0.68/-0.93)\\
1.0	&3.76 (+0.66/-1.09)&3.43 (+0.64/-1.01)&2.91 (+0.58/-0.88)\\
\enddata
\end{deluxetable}

\begin{figure}[bth]
\plotone{f1.eps}
\end{figure}

\begin{figure}[bth]
\plotone{f2.eps}
\end{figure}

\begin{figure}[bth]
\plotone{f3.eps}
\end{figure}

\begin{figure}[bth]
\plotone{f4.eps}
\end{figure}

\begin{figure}[bth]
\plotone{f5.eps}
\end{figure}

\begin{figure}[bth]
\plotone{f6.eps}
\end{figure}

\begin{figure}[bth]
\plotone{f7.eps}
\end{figure}

\begin{figure}[bth]
\plotone{f8.eps}
\end{figure}

\end{document}